\documentstyle[graphicx,12pt]{elsart} 

\begin{document}
\begin{frontmatter}

\title{Fragmentation path of excited nuclear systems}

\author[Lab]{M. Colonna},
\author[Fir]{G. Fabbri},
\author[Lab]{M. Di Toro},
\author[Fir]{F. Matera},
\author[Mon]{H.H. Wolter}

\address[Lab]{Laboratori Nazionali del Sud, 
        Via S. Sofia 44, I-95123 Catania,\\
and Physics-Astronomy Dept., University of Catania, Italy}

\address[Fir]{Dipartimento di Fisica, 
        Universita' di Firenze, and
        INFN, Sez.di Firenze, Italy}

\address[Mon]{Sektion Physik, Universitat Muenchen, Garching, Germany}



\begin{abstract}
We perform a study of the fragmentation path of excited nuclear sources, 
within the  framework of a stochastic mean-field approach. 
We consider the reaction $^{129}$Xe + $^{119}$Sn at two beam 
energies: 32 and 50 MeV/A, for central collisions. 
It is observed that, after the 
compression phase the system expands towards a dilute configuration
from which it may recontract or evolve into a bubble-like structure.
Then fragments are formed through the development of volume and/or
surface instabilities.  
The two possibilities co-exist at 32 MeV/A, leading to quite different
fragment partitions, while at 50 MeV/A the hollow configuration is observed 
in all events. Large variances are recovered 
in a way
fully consistent with the presence of spinodal decomposition remnants.  
Kinematical properties of fragments are discussed and suggested
as observables very sensitive to the dominant fragment production mechanism.  
A larger radial collective flow is observed at 50 MeV/A, in agreement with
experiments.   
\end{abstract}
\maketitle
\end{frontmatter}

\section{Introduction}

The observation of several intermediate mass fragments (IMF) in 
heavy ion collisions and the possible connection to the occurrence of
liquid-gas phase transitions in nuclear systems  
are still among the most challenging problems in heavy ion reaction
physics. 

Recently many efforts have been devoted to indentify  
relevant observables that  
may signal the occurrence of phase transitions in
finite nuclei \cite{Frankland,Moretto,Dagostino,Chomaz,EPJA}.
These studies are mostly based on the investigation of 
thermodynamical properties of finite systems at 
equilibrium and on critical behaviour analyses. 
In such a context,       
a description of the dynamics of fragment formation can provide
an important complementary piece of information to
learn about the path of the fragmentation process.    
Indeed the study of the dynamical evolution of complex excited nuclear
systems may shed some light on the relevant fragmentation mechanisms
and on density and temperature conditions where fragments are formed.  
This opens the possibility to get insight into 
the behaviour of the system in such regions, far from normal values, 
encountered 
along the fragmentation path, and hence on its equation 
of state (EOS). 
Eventually at the freeze-out
configuration one may check how the available phase space has 
been filled and compare results obtained for some
relevant degrees of freedom and physical observables to 
the values expected at thermodynamical equilibrium.    
At the same time    
one may try to identify some specific observables that keep the
memory of the fragmentation mechanism and are sensitive to the EOS. 

In this article we present an analysis of the fragmentation dynamics
of excited nuclear sources that are formed in central heavy ion collisions
at intermediate energies.  This study is performed within the framework
of stochastic mean field approaches \cite{TWINGO,BOB,Salvo,Matera}. 
As extensively discussed also in Ref.\cite{report},  
according to this theory,  
the fragmentation process is dominated by the growth of volume (spinodal)
and surface instabilities encountered during the expansion phase of the 
considered excited systems. 
Hence the fragmentation path is driven 
by the amplification of the unstable collective modes.  
The correct description of  
the degree of thermal agitation, that determines the amplitude of the
stochastic term incorporated in the treatment,  
is essential since fluctuations 
provide the seeds of fragment formation.

We notice here that 
this scenario for fragment formation occurs in several 
many-body systems. 
Indeed spinodal instabilities have been shown to drive the   
fragmentation path of dilute classical systems of particles
interacting through short-range forces, for which
exact dynamical calculations can be performed \cite{Bertrand,Pratt,Dorso}, 
and are a well known
mechanism in the decomposition of binary alloys \cite{Gunton}.    
The role of spinodal instabilities
in nuclear fragmentation has been recently investigated 
also in experimental data \cite{Borderie_PRL,EPJA}. 

Multifragmentation studies have been undertaken also in the framework
of molecular dynamics approaches for fermions \cite{feldi,Ono,Onishi}. 
It would be interesting to investigate whether, according to these
models, fragments appear already in the high-density phase, due
to the presence of strong N-body correlations, or
a composite almost uniform system may survive until it enters low
density regions, as it happens within our scenario. In the two cases
fragment formation would be sensitive to different regions of the nuclear EOS
and one expects to see differences expecially in charge and velocity
correlations and in the isotopic content.   

In this article, calculations are performed using the stochastic mean-field
method 
described in Ref.\cite{Salvo}.
With respect to previous studies, the method employed here has a 
larger domain of applicability.  
Indeed we implement  
density fluctuations locally in  
coordinate space to mimic 
the effect of the local thermal
equilibrium fluctuations. 
This can be done at each time and in each configuration,  
the only requirement being the condition of local thermal equilibrium. 
On the other hand, in the approaches used 
in previous works \cite{BOB}, a stochastic
force was introduced, whose strength can be tuned to reproduce 
the correct fluctuation amplitude of few
fragmentation modes, thus requiring the knowledge of the most important
ones. 

Moreover, the method of Ref.\cite{Salvo} can be easily extended to  
fragmentation studies of charge asymmetric systems \cite{Virgil2}, 
of large interest 
nowadays. In fact fluctuations can be 
implemented separately for protons and neutrons, on the basis of their 
respective kinetic equilibrium fluctuations.

We perform calculations for systems of experimental interest, 
showing how physical observables can be related to the degree and
the kind of instability encountered.


The paper is organized as follows:  In Section 2 we recall the main 
features of the method, describing in more detail the procedure to
inject fluctuations; In Section 3 results on the fragmentation
dynamics of excited nuclear sources are presented.  Some 
conclusions are drawn in Section 4.

\section{Description of the method}

 
Theoretically the evolution of complex systems  
under the influence of fluctuations can be described
by a transport equation with a stochastic fluctuating term, the so-called
Boltzmann-Langevin equation (BLE):
\begin{equation}
{{df}\over{dt}} = {{\partial f}\over{\partial t}} + \{f,H\} = I_{coll}[f] 
+ \delta I[f],
\end{equation}
 where $f({\bf r},{\bf p},t)$ is the one-body distribution function, 
 $H({\bf r},{\bf p},t)$ is the one-body Halmitonian and 
$\delta I[f]$ represents the stochastic part of the two-body
collision integral \cite{Ayik,Randrup}.
Exact numerical solutions of the BLE are very difficult to be obtained and
they have only been calculated for schematic models in two 
dimensions \cite{Fiorella}. 
Therefore various approximate treatments of the BLE have been introduced
\cite{Eric,BOB,Salvo}. 
In the Brownian One Body (BOB) dynamics, developed in Ref.\cite{BOB}, 
the fluctuating part of the collision integral $\delta I$ is replaced by
a stochastic force added to the standard Botzmann-Nordheim-Vlasov (BNV) 
equation 
(i.e. Eq.(1) without
stochastic term), 
the strength of
which can be tuned to correctly describe the growth of the most important 
unstable modes. 

Here we will follow the method introduced in Ref.\cite{Salvo}.
 
Within the assumption of local thermal equilibrium, the stochastic term 
of Eq.(1) essentially builds the kinetic equilibrium fluctuations typical
of a Fermi gas: $\sigma_f^2 = f(1-f)$, where $f({\bf r},{\bf p},t)$ 
can be approximated by a Fermi-Dirac
distribution, $f({\bf r},{\bf p},t) = 1/(1 + e^{(\epsilon - \mu)/T})$ 
with local chemical potential and temperature, 
$\mu({\bf r},t)$, $T({\bf r},t)$.
We project on density fluctuations obtaining:
\begin{equation}
{\sigma^2_{\rho,eq}}({\bf r},t) = {1\over V}\int {{d{\bf p}}\over{h^3/4}}
\sigma^2_f({\bf r},{\bf p},t) = {T \over V} {{3\rho}\over{2\epsilon_F}}
(1-{{\pi^2}\over{12}} ({T \over {\epsilon_F}})^2 + ...),
\end{equation}
where we have used the Sommerfeld expansion around $\epsilon = \mu$ for
small $T/\epsilon_F$ ($\epsilon_F$ denotes the Fermi energy). 

Actually this kinetic equilibrium value is reached asymptotically for an
ideal gas of fermions. 
After a small time step $\Delta t$, the value reached for the density
fluctuations can be approximated by:
\begin{equation} 
\sigma^2_\rho({\bf r},t,\Delta t) = \sigma_{\rho,eq}^2({\bf r},t)
{{2\Delta t}\over{\tau_{coll}({\bf r},t)}},
\end{equation}
where $\tau_{coll}$ is the damping time associated with the two-body 
collision process.  
At temperatures around 4 MeV, that are typically reached in fragmentation
processes, 
$\tau_{coll}$ is of the order of 50 fm/c \cite{Jorgen}. 
In our calculations fluctuations are injected each $\Delta t = 5~ fm/c$
and their amplitude is scaled accordingly (see Eq.(3)).
In the cell of ${\bf r}$ space being considered, the density fluctuation
$\delta\rho$ is selected randomly according to the gaussian distribution
$exp(-\delta\rho^2/2\sigma_{\rho}^2)$. This determines the variation of the
number of particles contained in the cell. A few left-over particles are 
randomly distributed again to ensure the conservation of mass. 
Momenta of all particles are finally slightly shifted to ensure momentum
and energy conservation.
\begin{figure}[htb]
\includegraphics*[scale=0.55]{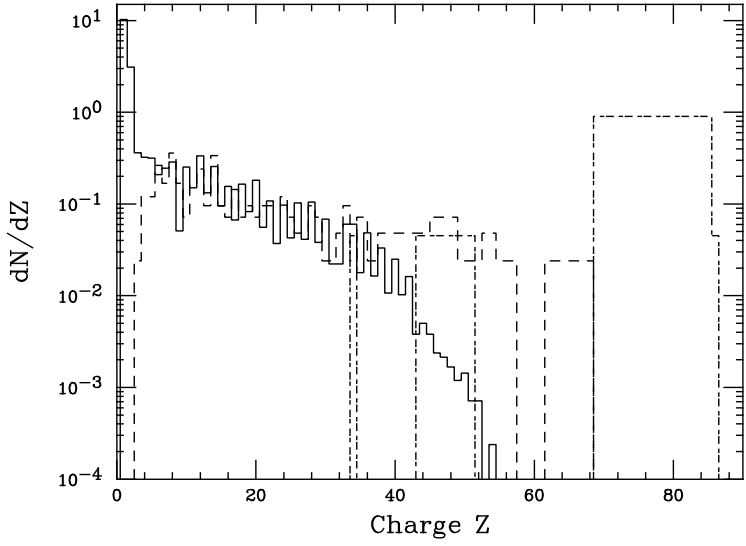}
\caption{\it 
Primary charge distribution obtained for the nuclear source: 
A = 247, Z = 103, T = 8.3 MeV, R = 6.6 fm, $\beta$ = 0.09 r/R, 
with (long-dashed histogram) and without (short-dashed histogram) including
fluctuations.  
The full histogram represents the final charge distributions obtained
in the case including fluctuations.} 
\end{figure}   

The fluctuations introduced are then amplified by the unstable mean-field. 
It is important to notice that the characteristic growth time 
for the unstable modes is smaller ($\tau\approx 30~fm/c$) \cite{report}
than the collisional damping time, so two-body 
collisions act just as a seed to perturbe the density profile of 
the system, but the dynamics is essentialy driven by the 
propagation of the mean-field instabilities.  

\begin{figure}[htb]
\includegraphics*[scale=0.55]{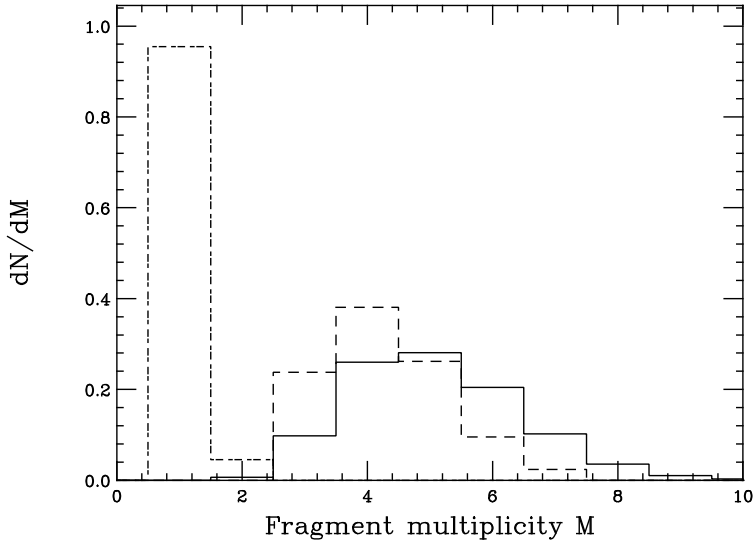}
\caption{\it 
Fragment multiplicity distribution obtained for the 
same case as in Fig.1.  
Symbols are as in Fig.1.} 
\end{figure}

\section{Presentation of the results and discussion}

In order to investigate  the fragmentation path followed
by excited dilute sources,  
we start considering a spherical system of radius $R = 6.6~fm$ containing 
A = 247 nucleons and  Z = 103 protons.
 T = 8.3 MeV is the initial temperature and the 
system undergoes a self-similar expansion with velocity
$\beta(r)=v(r)/c = 0.09~r/R $.  
These 
properties correspond to the composite source formed in the reaction
 $^{129}$Xe + $^{119}$Sn at 32 MeV/A, at zero impact parameter, just after 
the maximum compression has been reached,
according to deterministic BNV calculations
(no fluctuations included). 
For the same system calculations have been performed also using 
the BOB treatment \cite{BOB}.  

We use the test particle method, with 200 test particles per nucleon.
This ensures a quite good mapping of the phase space, since each test particle
is associated with a triangular function of width $\approx 1~fm$
\cite{TWINGO}.
We adopt a Skyrme-like parameterization of the mean-field potential,
corresponding to a soft EOS (compressibility modulus K = 200 MeV). 

The dynamical evolution of the system is followed until 240 fm/c.
Fluctuations are introduced since 
the beginning, then they 
are already present when the system encounters instabilities.  

To underline the important role of fluctuations in the dynamical evolution of 
the system,
we also perform simulations turning off fluctuations, 
i.e. following the standard BNV dynamics. 
In this case, the initial    
spherical symmetry and uniform radial distribution
are so well preserved that the system, after a moderate
expansion,  eventually recontracts and does not fragment while in the full
treatment (including fluctuations) fragmentation is observed.

In Fig.1 we show the charge distribution obtained at the end of the 
dynamical simulations, with (long-dashed histogram) and 
without (short-dashed histogram)
fluctuations.  The final charge distribution, obtained after secondary
de-excitation has been taken into account \cite{Simon}, is shown 
for the simulations including fluctuations (full histogram).
We remark that  
the final charge distribution is in good agreement with the results
obtained with the BOB method \cite{BOB} 
and with experimental data \cite{John}. 

   
It is clear that in the case without fluctuations the system essentially
recontracts, as signaled by the peak at very large Z.
This is evident also from the fragment ($Z\geq 3$) multiplicity distribution 
(Fig.2). On the other hand, in the calculations
including fluctuations the system breaks up into 5 fragments, on average.

\begin{figure}[htb]
\includegraphics*[scale=0.7]{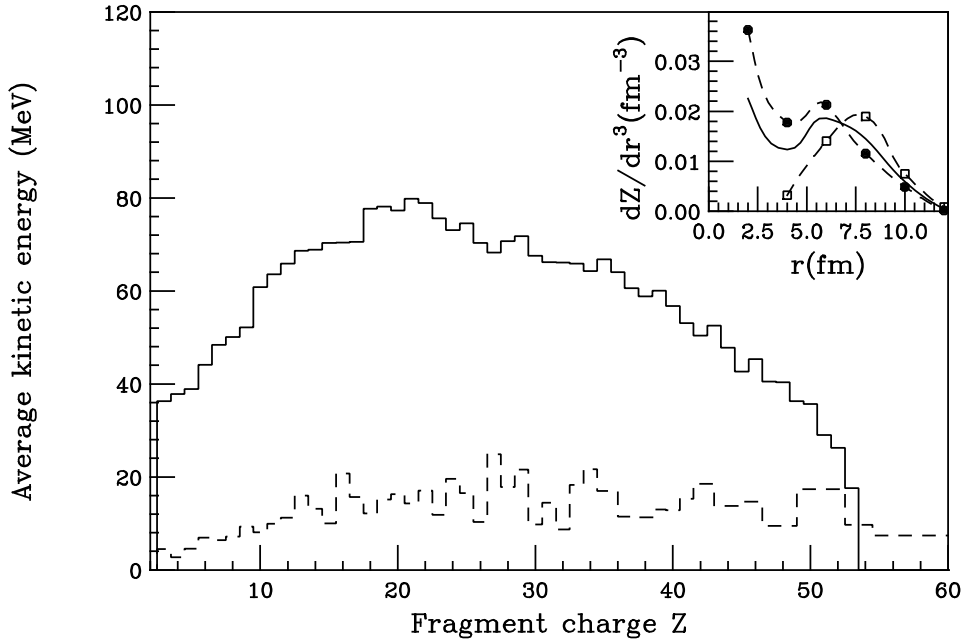}
\caption{\it 
Average kinetic energies for 
primary (dashed histogram) and final (full histogram) fragments 
obtained in the same case as in Fig.s 1,2. 
In the inset we show the fragment charge density as a function of the radial 
distance r (full line). The results obtained considering only multiplicities
M=3,4 (full circles) and M=5,6,7 (squares) are also displayed.} 
\end{figure}
In fig.3 
we show the average kinetic energy, as a function of the fragment 
charge, obtained for primary (dashed histogram) and final 
(full histogram) fragments.
\begin{figure}[htb]
\includegraphics*[scale=0.7]{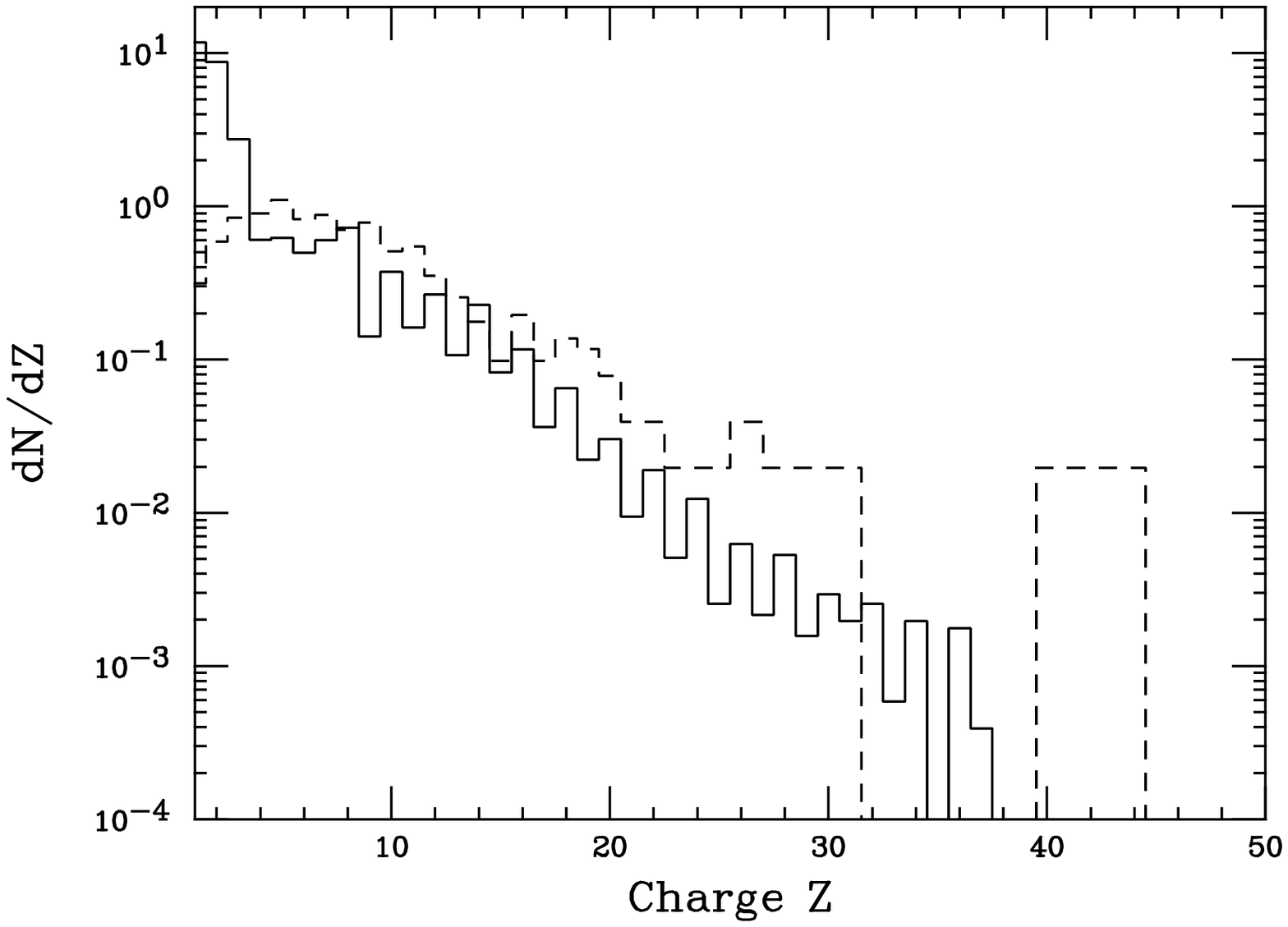}
\caption{\it 
Charge distribution obtained for the nuclear source: 
A = 244, Z = 101, T = 10 MeV, R = 7.5 fm, $\beta = 
0.13~ r/R$,  
for final (solid histogram) and primary (dashed histogram) 
fragments.}
\end{figure} 
The fragment velocities are essentially due to the (low) 
radial collective flow, 
related
to the expansion of the system, and to the Coulomb repulsion.
The latter is responsible of the large difference observed between primary
and final kinetic energies. Indeed at the freeze-out
configuration the Coulomb repulsion is still important.
\begin{figure}[htb]
\includegraphics*[scale=0.7]{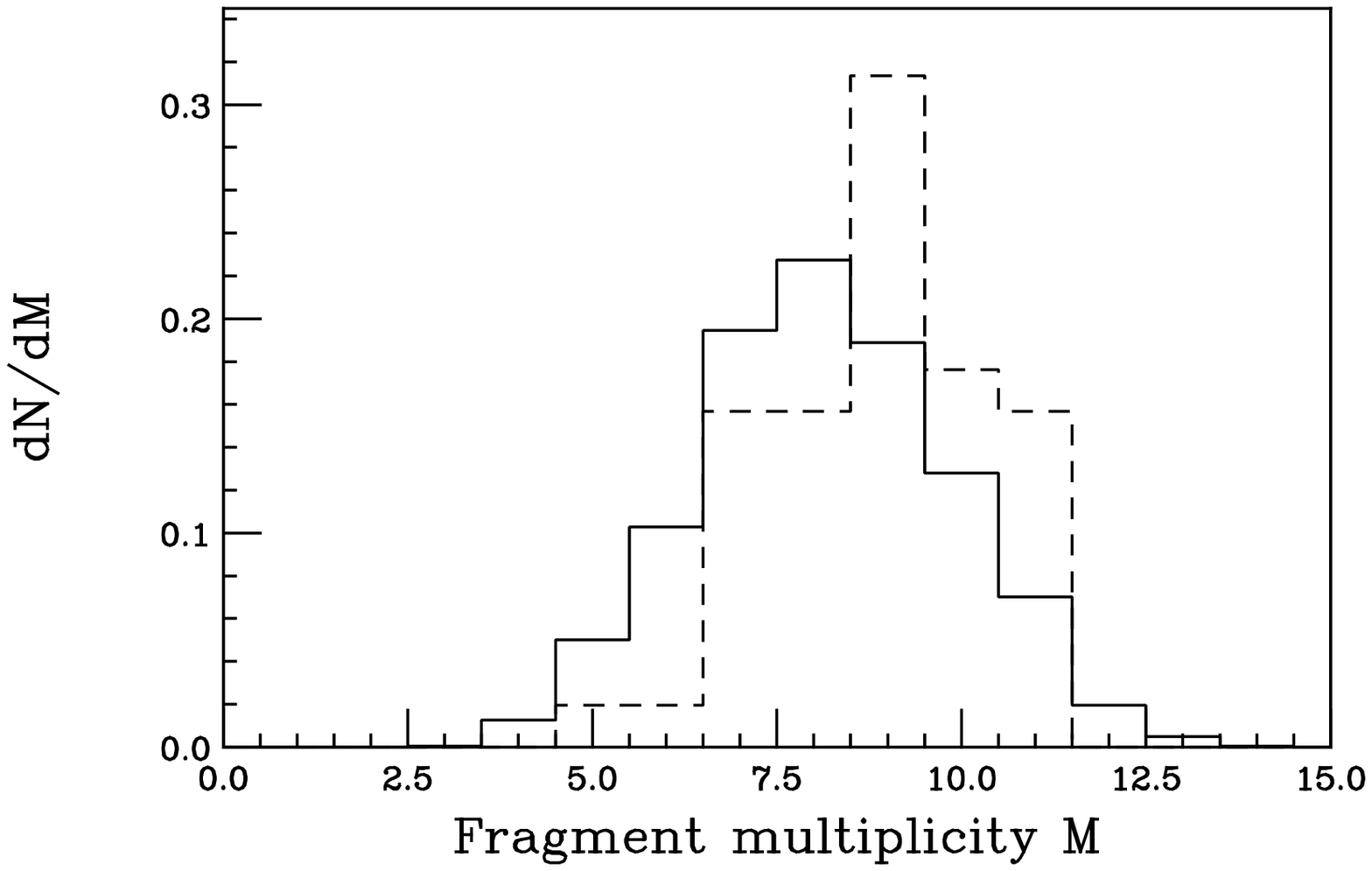}
\caption{\it 
Multiplicity distribution for the same case as in Fig.4,
obtained for final  
(solid histogram) and primary (dashed histogram) fragments.}
\end{figure} 

\begin{figure}[htb]
\includegraphics*[scale=0.7]{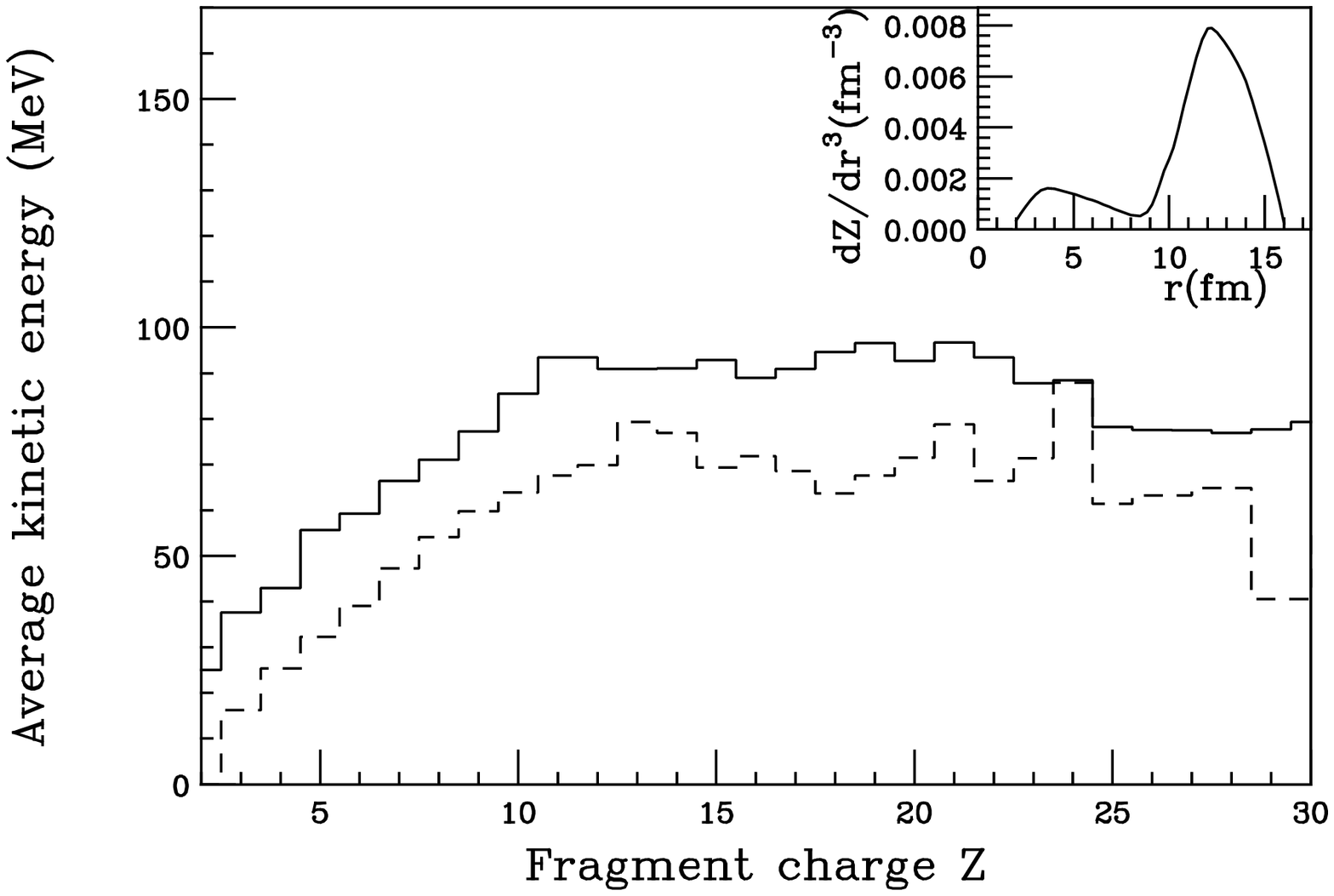}
\caption{\it 
Fragment average kinetic energy 
as obtained in our simulations 
for primary (dashed histogram) and final (full histogram) fragments.
In the inset we show the fragment charge density as a function of the 
radial distance r.}
\end{figure}  
The structure of the freeze-out configuration is illustrated in the inset, 
where we plot the charge density as a function of the distance from
the center of mass of the system, $r$ (full line).
It is possible to observe that, on average, 
the charge of the fragmenting system  
is located
close to the center of mass and at 
a distance $r\approx 7~fm$, where a second bump is observed.  
This interesting behaviour is due to the fact that, 
when fluctuations are introduced, the
system may recontract in some cases, leading to the formation of
fragments close to the center of mass, or develop into a 
bubble-like structure.  Of course, events where some fragments 
form close to the center, accompanied by lighter fragments located at larger
distance are also observed.    
From a more detailed analysis, it is possible to see that in events with
low IMF multiplicity, $M_{IMF} = 3,4$, the charge 
is mostly concentrated close to the center of mass (full circles),
indicating that more compact configurations with coupled surface and volume
instabilities \cite{RPA} are formed,  
while in events with larger multiplicity, from 5 to 7, fragments
are observed around $r\approx 8~fm$, at the boundary of a 
hollow configuration (squares).  
The growth times of instabilities are different in the two cases,  
the hollow configuration being more unstable. In fact the system reaches
a higher degree of dilution in this situation.   
Hence two scenarios of fragment formation seem to co-exist, yielding
quite different fragment partitions and a broad charge distribution
(see Fig.1).
The fact that large IMF's are mostly located close to the center 
is reflected also in the decreasing behaviour of the average kinetic energy
observed at large Z (see Fig.3).   

\begin{figure}[htb]
\includegraphics*[scale=0.7]{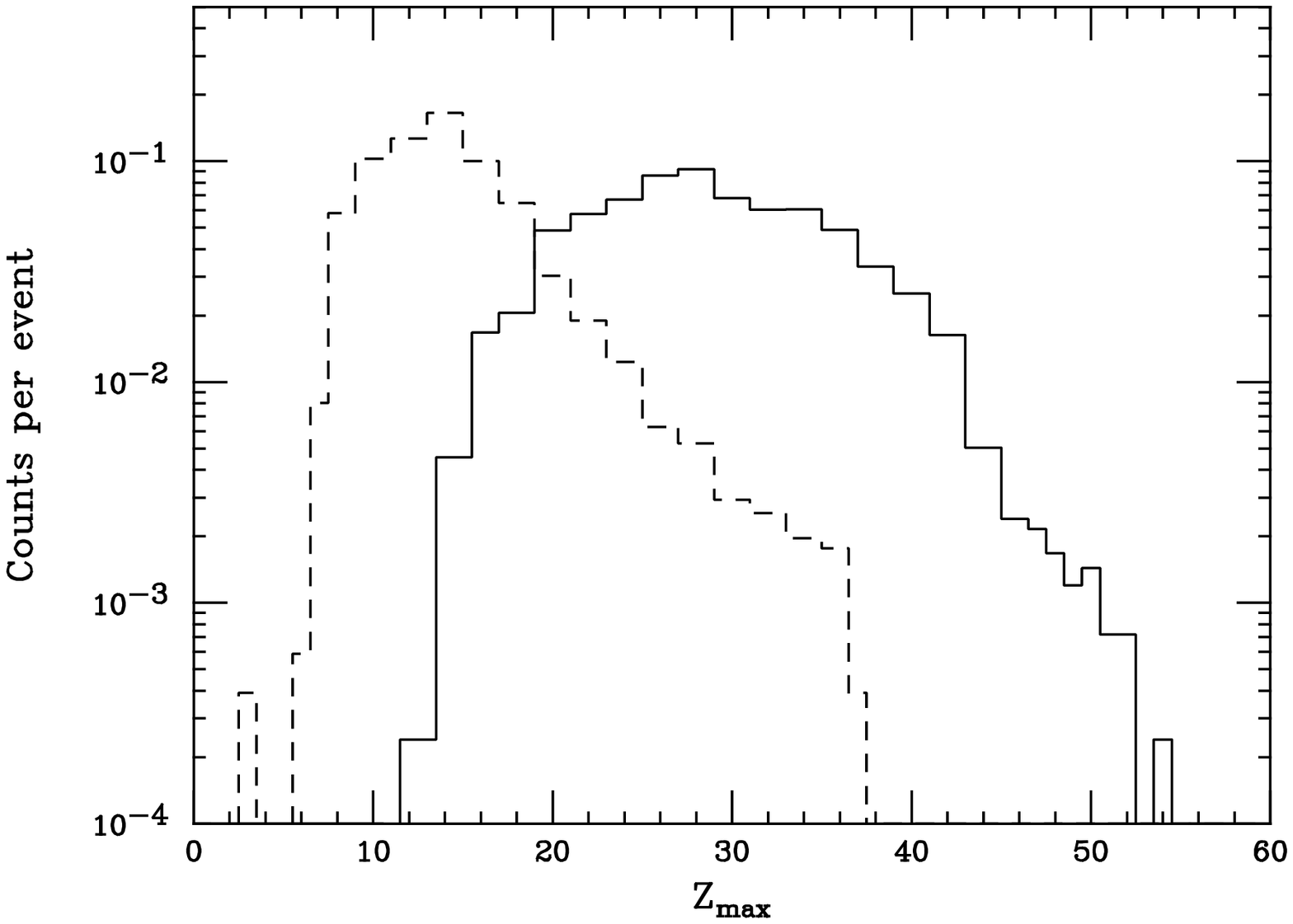}
\caption{\it 
Distribution of the largest fragment obtained in the reaction  $^{129}$Xe + $^{119}$Sn at
32 MeV/A (full histogram) and 50 MeV/A (dashed histogram).}
\end{figure} 
To appreciate how the fragmentation mechanism evolves with the 
energy stored in the system, now 
we turn to consider 
the dynamical evolution of an excited spherical source
with the following properties: 
A = 244, Z = 101, temperature T = 10 MeV, 
radius R = 7.5 fm, and with a self similar radial velocity $\beta = 
0.13~ r/R$. 
This system is obtained in 
the collision  $^{129}$Xe + $^{119}$Sn at 50 MeV/A, 
at the beginning of the expansion  
towards densities lower than the normal value. 

In this case 
it is observed that, as a property of the average dynamics, i.e.
even without including fluctuations, the nucleus expands into a hollow,
quasi stationary configuration, that evolves rather slowly,
while, as we have seen before, at lower energies  
the system essentially recontracts in absence of fluctuations 
\cite{Batko}.     

Actually 
the spatial geometry of the fragmenting system has been the object
of several investigations, within BUU-like approaches \cite{borderie}.
In many cases the formation
of a bubble-like structure has been reported.
This is due to the competition
between the radial expansion flow and the velocity field that develops
at the surface. In fact the latter tries to recompact the system. 

The hollow configuration is unstable
against density fluctuations that break
the spherical symmetry. 
Hence,
including self-consistent fluctuations in the dynamics, 
fragment formation is observed.
Fig.4 shows the fragment charge distribution (dashed histogram) obtained 
at the freeze-out time,
t = 140 fm/c. The final charge distribution is 
also displayed (full histogram).    

The system explodes into lighter IMF's, as indicated already by the primary
charge distribution. Moreover 
the number of nucleons and light
particles that are emitted while fragments are formed is larger in this
reaction, with respect to the case at 32 MeV/A. 
In fact, the total charge of the IMF's ($Z\geq 3$) produced at the freeze-out 
changes from 84 (at 32 MeV/A) to 75 (at 50 MeV/A). 
 
The IMF multiplicity distribution is presented in Fig.5. 
It is interesting to notice that the average number of primary IMF's is larger
than the final one. This is due to the fact that 
very light IMF's (Z=3 or 4) present at the freeze-out configuration
may decay into free nucleons and light particles.     

The spatial distribution of fragments is illustrated in the inset of Fig.6.
Now it is possible to observe that fragments are mostly distributed 
around $r = 12~fm$. 
So, with respect to the case at lower energy previously discussed,  
the bulk of the system appears more concerned by the instabilities 
and the system breaks up into a larger number of smaller fragments. 
The average kinetic energy, as a function of the fragment charge, 
is shown in Fig.6, for primary (dashed histogram) and final fragments 
(full histogram).  
Now primary and final kinetic energies are similar 
since at the freeze-out configuration fragments are already quite well 
separated and the Coulomb
repulsion is small. We notice also the larger radial collective flow
obtained in this case.

Finally, in Fig.7 we report 
the distribution of the largest final fragment, $Z_{max}$ 
observed event by event, for the two reactions considered.  
A broader distribution is obtained at 32 MeV/A, indicating that 
configurations where large fragments are present co-exist with cases
where only light IMF's are formed.    
On the other hand, large fragments ($Z>40$) are not observed at 50 MeV/A,
fragment partitions are less fluctuating event by event and the
$Z_{max}$ distribution has a smaller width.  

We would like to notice that the results reported here are in qualitative
agreement with the trend observed in 
experimental data \cite{John,sylvie}.  
For the reaction at 32 MeV/A, a careful comparison between data and
stochastic calculations performed with BOB can be found in Ref.\cite{John}.
For the same reaction the presence of spinodal decomposition remnants,
i.e. the observation of a few events with nearly equal-sized fragments,
has been found in the data, as well as in the simulations \cite{EPJA}.  
    
It would be interesting to try to extract information, from the experimental
point of view, on the shape of fragment configurations, such as, for instance, 
the formation of a bubble-like structure that we observe in the simulations
at 50 MeV/A. 
This kind of analysis has been recently undertaken also in light ion
induced fragmentation studies \cite{Cugnon,Karna}.  




\section{Conclusions}

In this article we perform a study of the fragmentation path followed 
by excited nuclear systems formed in central heavy ion collisions,
at two different beam energies. According to stochastic 
mean-field theories,   
fragmentation happens during the expansion phase and thus it is related to
the properties of the nuclear system at low density and, in particular, 
to the occurrence of volume and shape instabilities. This opens the 
possibility to learn about the behaviour of the nuclear EOS at low density. 

Mean-field calculations show that, 
in central heavy ion collisions,  
the reaction mechanism evolves, with increasing beam energy, 
from the formation
of a single primary source (i.e. incomplete fusion) 
to fragmentation. 
In particular, in a given energy range, 
monopole oscillations push the system
towards a metastable configuration, that may recontract, in some cases, 
or develop into a hollow configuration that eventually fragments. 
This is what is observed, for the reaction  $^{129}$Xe + $^{119}$Sn, at 32 MeV/A.    
Large fluctuations in the fragment configurations appear in this case,
because of the coexistence of the two mechanisms.  This is in line with
recent experimental observations, that have been related to the occurrence 
of negative specific heat \cite{nicolas}. 
The presence of a
few events with equal-sized fragments, that may signal the occurrence of
spinodal decomposition,  
has also been reported \cite{EPJA}. 
At 50 Mev/A the dynamics is dominated by the formation of hollow 
structures that break up into fragments. 
A larger collective flow is observed in this case and a larger number 
of IMF's is formed already at the primary level, together with light 
particles. 
In this way we expect to see the transition from the fragmentation 
regime to vaporization, observed at higher energies.  

\end{document}